# Copper Electrodeposition for 3D Integration

Rozalia Beica, Charles Sharbono, Tom Ritzdorf
Semitool, Inc.
655 West Reserve Drive
Kalispell, MT 59901 USA

*Abstract*-Two dimensional (2D) integration has been the traditional approach for IC integration. Increasing demands for providing electronic devices with superior performance and functionality in more efficient and compact packages has driven the semiconductor industry to develop more advanced packaging technologies.

Three-dimensional (3D) approaches address both miniaturization and integration required for advanced and portable electronic products. Vertical integration proved to be essential in achieving a greater integration flexibility of disparate technologies, resulting in a general trend of transition from 2D to 3D integration in the industry.

3D chip integration using through silicon via (TSV) copper is considered one of the most advanced technologies among all different types of 3D packaging technologies.

Copper electrodeposition is one of technologies that enable the formation of TSV structures. Because of its well-known application for copper damascene interconnects, it was believed that its transfer to filling TSV vias would be easily adopted. However, as any new technology at its beginning, there are several challenges that need to be addressed and resolved before becoming a fully mature technology.

This paper will address the TSV fill processes using copper electrodeposition, the advantages as well as difficulties associated with this technology and approaches taken to overcome them. The effect of wafer design on process, including necessary process optimization that is required for achieving void-free filling will be discussed.

## I. INTRODUCTION

Two-dimensional approaches have been traditionally applied for IC integration. Continuous pressure for new IC packages that can address the ever-expanding consumer electronics market demands for increased functionality and performance while reducing the size and cost, has driven the semiconductor industry to develop more innovative packaging, using vertical, 3D integration [1-4].

3D packaging has been extensively studied in recent years, in particular for advanced mobile devices. 3D stack packages were initially developed using wire bonding for memory packages and SDRAM of cellular phones [2].

General advantages of 3D packaging technologies include form factor miniaturization (size and weight), integration of heterogeneous technologies in a single package, replacement of long 2D interconnects with short vertical interconnects, and the reduction of parasitics and power consumption. Therefore, 3D packaging can offer significant advantages in performance functionality and form factor for future technologies.

Based on the stacking method, there are different type of 3D packages, which include on-chip 3D integration based on layer-by-layer build-up of functional layers within a chip; 3D stacking with die-to-die stacking or package-to-package stacking (package-on-package POP or package-in-package PIP); and 3D ICs (3D integration of ICs), which have die-to-die interconnection with through-silicon vias.

The typical trend in technology development continues to move from 2D configuration to 3D stacking (with wire, bumps and micro vias), and then to 3D ICs with TSV interconnects in order to reduce form factor, increase silicon efficiency, and have shorter interconnects [5].

Wire bonding today, even if it has the advantage of being a mature, well-known and characterized technology, is limited when it comes to 3D chip stacking in terms of density and performance. To accommodate wire bonds both vertically and horizontally, higher density packages are needed; therefore miniaturization of the devices is more difficult to achieve. The I/O density is also reduced, and the connections are long and limited to the periphery of the device.

Conventional flip-chip connections have their size limitations as well, which is why this technology cannot be broadly adopted for chip stacking.

By providing shorter connections with TSV, the distance of information flow on a chip will be significantly reduced (by up to 1000x). TSV connections also allow the addition of increased number of channels or pathways (up to 100x more than 2D chips), necessary for the information to flow. TSV can replace less-efficient wire bonds for transferring signals off the chip, increase speed while reducing power consumption and enables uniform power delivery to all parts of the device [6]; it makes possible the formation of higher density and higher aspect ratio connections, allowing the integration of multi-chip systems entirely within the silicon with a better packing density than traditional 3D packaging methods. Because of its advantages over traditional technologies, the TSV approach, although still expensive and difficult to implement in high volume for most applications, continues to have a growing interest, being one of the hottest topics of the semiconductor industry today, not only for 3D packaging purposes, but also for the raw benefits offered in a single device, such as higher frequency [7].

## II. PROCESS FLOW FOR TSV CHIP STACKING

Depending on the application, there are two approaches adopted for TSV chip stacking: blind-via filling before wafer thinning and through-via filling after wafer thinning [8-10].

The vias in the first approach (as presented in Figure 1) are filled with copper from the front side of the silicon wafers, very similar to the typical interconnect applications. After the vias are filled and overburden is removed, the wafer is thinned at the backside of the wafer/vias until the TSV electrodes are exposed.

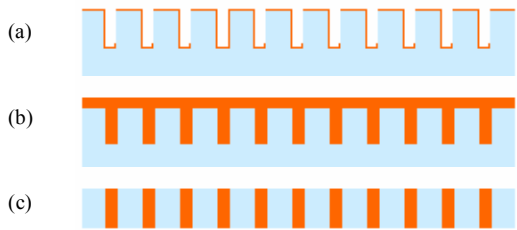

Fig. 1. Blind-via filling before wafer thinning [10]: (a) Via formation, insulator/barrier/seed deposition; (b) Via filling; (c) Metal removal, wafer thinning

The second approach (as presented in Figure 2) is to fill vias through via holes after wafer thinning. In this case the seed layer is deposited on one side of the wafer.

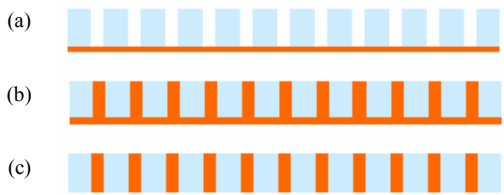

Fig. 2. Through-via filling after wafer thinning [10]: (a) Thinning / via formation/seed deposition; (b) Via filling; (c) Seed removal

Due to the difficulties in handling and plating thinned wafers, the first approach has gained more popularity becoming the most preferred TSV stacking method in the industry today.

## III. MATERIALS AND PROCESSES FOR TSV ELECTRODES

There are different materials and methods that have been proposed and tested for TSV applications. Feature dimensions, final deposit characteristics and process reliability, along with cost considerations are the main limiting factors of applicability of these materials and methods.

The most common material used for via filling and formation of TSV electrodes is copper, due to its properties and compatibility to conventional multilayer interconnection in FEOL and back-end-of-line (BEOL) processes [10]. Alternative conductive materials include chemical vapor deposited (CVD) tungsten and doped polysilicon. Both of these materials provide a better match of the coefficient of thermal expansion (CTE) with silicon, very important especially for larger features, however, at the expense of lower electrical and thermal conductivity than copper. Although announcements of using all these materials have already been made, which one is going to gain more market acceptance remains to be seen [7,10-12].

Copper can be applied through CVD, plasma vapor deposition (PVD) and most commonly through electrodeposition. The CVD method, although a well-known technology from the front-end-of-line (FEOL) contact via applications, is more suitable for smaller features (up to 2μm in diameter). Electrodeposition processes are typically performed at room temperature and pressure, so the equipment is less complex and expensive than vacuum deposition equipment. Because of the high costs involved with CVD applications and the use of unstable and hazardous organo-metallics, which result in deposits that have lower purity and higher resistivity, as well as the advantages that come with electrodeposition processes, such as faster speed, higher stability, ease of control and maintainability, most TSV developments have been focusing on using electrodeposition as the preferred application method [7].

The formation of TSV electrodes using copper deposition, following the via first stacking approach described in Figure 1, starts with via etching in the silicon (deep reactive ion etch or laser technologies), followed by a deposition through plasma enhanced chemical vapor deposition (PECVD) of an insulator ($SiN_x$ or $SiO_2$). The insulator is then covered with barrier layers such as TiN or TaN using metal organic chemical vapor deposition (MOCVD) or PVD technology and subsequently with a conductive Cu seed layer, either through CVD or PVD, depending on the features shape and aspect ratio. Once the seed layer is present inside the features it provides the necessary conductive layer required for electrodepositing the TSV structures [10].

Depending on via sizes and applications, there are typically three types of deep via copper (as presented in Figure 3): lining and full filling with and without stud [4,10].

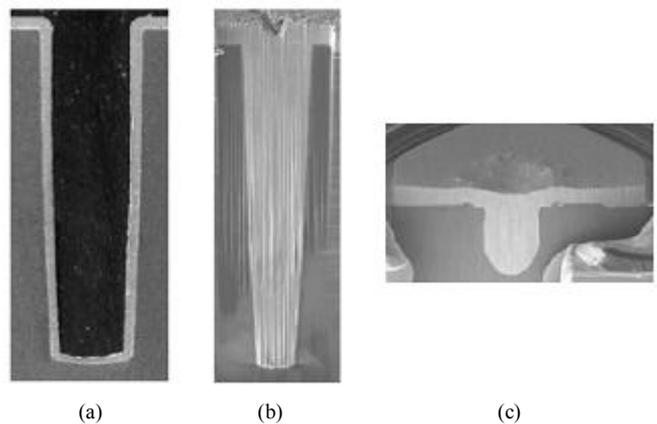

Fig. 3. Types of deep-via deposition:
(a) lining; (b) full-filling, and (c) full-filling with stud formation

If via lining is more practical for larger features, more commonly found in MEMS and sensor applications, full filling is applied over a wider range of via dimensions, ranging from near damascene applications to larger vias used for sensor applications. Photoresist patterns can also be added to enable formation of a stud or additional connections, such as redistribution lines (RDL), for solder bonding.

After electrodeposition, the wafer is thinned and additional processing steps are applied to form the final 3D package.

## IV. Factors Affecting Via Filling Process Performance

Copper electrodeposition has been already applied for many years in the semiconductor and electronics industry, therefore the transfer of this technology, from applications such as copper damascene as well as printed wiring boards (PWB), to TSV applications, was believed to be easily done. That was not the case. Many of the conventional copper chemistries have been tried; however the results were not satisfactory. Defects such as seams, voids and inclusions of the electrolyte, which could have a negative impact on reliability by disrupting the ability to carry the electrical signal, occurred. There was need for better performing chemistries and processes able to address the limitations seen with the conventional materials and methods.

Copper, due to enhanced transport at the top of the via and a higher field at the small curvature – highly accessible areas) is going to build up faster, pinching-off and creating a voided deposit [13]. In order to achieve void-free filling, the deposition at the top needs to be significantly suppressed while the deposition at the bottom accelerated, to achieve a so-called "bottom-up" deposition. The composition of the copper electrolyte was found to have a significant effect on the process performance, with respect to quality of the deposit and maximum applicable deposition rates. Depending on bath composition and process parameters, different filling profiles can be obtained, as shown in Figure 4 [4, 10].

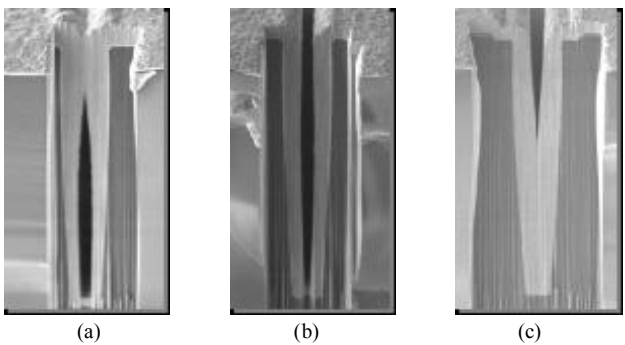

(a) (b) (c)

Fig. 4. Filling profiles: (a) sub-conformal profile, with void at the bottom; (b) conformal profile, with seam void; (c) conformal profile, void-free, bottom-up filling

By selection of the right organics, optimization of the chemistry in combination with the process parameters and reactor design, reliable TSV structures through copper electrodeposition can be achieved. Today's commercially available chemistries consist of three organic components designed as accelerator (or brightener), suppressor (or carrier) and leveler, in addition to the inorganic components of metallic ions, acid (either sulfuric or methanesulfonic based) and halide ions (chloride being the most commonly used). Additionally, very important characteristics of the electrolytes necessary for achieving void-free filling are good wettability (especially for high aspect ratio features) and stability (for larger features that require longer processing times).

Besides the need for advanced, high performing chemicals, wafer design (via profile and smoothness), seed layer continuity and high performance equipment are equally critical factors needed in order to successfully process TSV structures.

## V. Experimental

Experiments were performed using whole wafers (150-200mm) and wafer coupons (2x2cm$^2$). The test vehicles used varied substantially with respect to via profile, sizes and aspect ratios, with features ranging from 3-150μm in diameter and 10-250μm in depth, and aspect ratios of 2:1 to more aggressive ones, such as 10:1. All the wafers were plated using Semitool fountain type plating equipment. All the electrodeposition tests performed were done prior to wafer thinning (via first approach). The flow rate of the solution was kept constant at 5 gallons per minute (gpm) and the bath temperature was fixed at 25°C. The effects of the critical factors mentioned in the previous section, as well as of the various process parameters such as wafer rotation speed, current density, and waveform on process performance were studied. For process performance characterization, a focused ion beam (FIB, FEI dual beam 820) was used.

## VI. Results and Discussions

### A. The effect of via profile

Via profile, as mentioned in the previous section, was found to impact successful deposition in TSV structures. Depending on the process used to etch vias into the silicon, the different profiles, such the ones illustrated in Figure 5, can be obtained. These types of profiles could be *tapered* (usually with a profile angle between 85-90°), *straight* (with a profile angle of 90°) and *re-entrant* (smaller width at the top of the via compared to the center and bottom of the via).

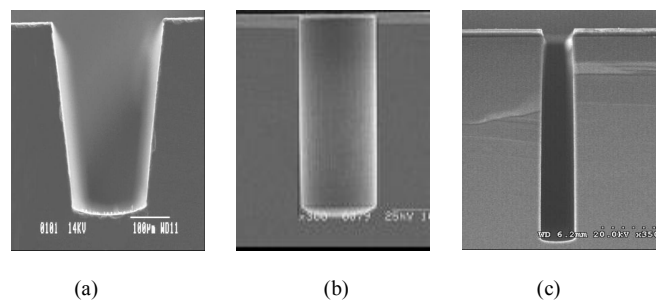

(a) (b) (c)

Fig. 5. Via profiles: (a) tapered; (b) straight; (c) re-entrant
(Source: Alcatel Micro Machining Systems)

Tapered vias, compared with straight and re-entrant vias, ensure adequate sidewall seed coverage and ease via filling [14]. The larger the opening of vias, the easier it is to fill. In comparison, the re-entrant profiles were found to be the most difficult ones to work with, due to higher probability of non-uniform seed layer coverage, especially at the bottom of the feature, and the need of higher performing chemistries that have the ability to provide an increased suppression at the via top and, at the same time, highly accelerated deposition at the bottom.

Tapered profiles also ensure growth minimization in silicon die size and minimize the impact of wafer thinning on device performance and reliability [1,14].

### B. The effect of insulator/barrier/seed layer coverage

The deposition quality (sufficient thickness, uniform coverage and good interlayer adhesion) is very important. Depending on the method used for depositing these layers, as well as via profile and aspect ratio of the features, the characteristics of these layers could vary. While PECVD used for depositing oxide can be conformal, PVD processes for barrier and seed layers have the difficulty of forming continuous layers, especially for small vias with high aspect ratios [1]. Figure 6 shows a bottom void obtained during via filling, due to non-coverage with seed layer of the via bottom.

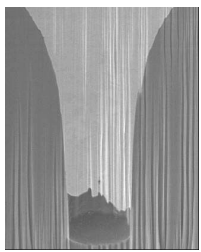

Fig. 6. Bottom via due to insufficient seed layer coverage

Seed layer enhancement or wet chemical deposition of seed layer, originally designed for damascene interconnect [4, 15, 16] has been demonstrated to be effective for deep 3D vias in overcoming the problems caused by insufficient coverage of seed layers [4].

### C. Wettability of the features

Wetting and ability to super-conformally fill vias of various sizes and aspect ratios, at faster speeds, are the main characteristics required with respect to the chemistries used for electrodeposition.

Chemistries with poor wetting capabilities will result in bottom voids, as shown in Figure 7a, especially for smaller vias with higher aspect ratios. Besides chemistry, oxides present on the surface of the seed layer could also cause poor wettability of the features (Figure 7b).

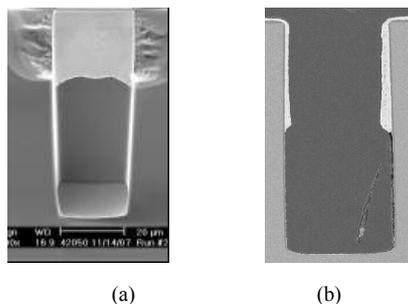

(a)  (b)
Fig. 7. Poor deep-via electrodeposition due to: (a) poor wetting; (b) oxides present on the seed layer (or inconsistent seed layer coverage)

Elimination of the wetting problems can be achieved by more aggressive pretreatments (impingement spray of liquid as prewet process), addition of a surfactant with high wetting capabilities either in the prewet step or in the deposition chemistries. The wetting agent needs to be chosen in such a way that it will not negatively affect via filling deposition.

In case the wetting defects are due to oxidized seed layer, an etching step with diluted acid solution can be applied.

### D. The effect of process parameters

Most process parameters were found to influence the deposit and filling mechanism within via to some extent. In our study, current density, waveform and bath composition were the parameters that had the most significant effect on the process performance.

With respect to *current density*, the probability of void formation becomes larger with increasing the current density, as shown in Figure 8. This is believed to be caused by increased current crowding at the via bottom and more significant mass transport limitations towards the bottom of via, when plating at higher current densities.

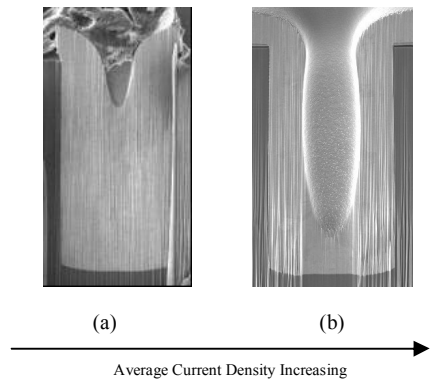

(a)  (b)

Average Current Density Increasing

Fig. 8. The effect of average current density on fill performance:
(a) full filling (12μm diameter/100μm depth); (b) partial filling (40μm diameter/100μm deep vias)

Improvement to the deposition process can be achieved by applying the proper *waveform* parameters. Pulse reverse deposition involves application of a forward current followed by a short, high-energy reverse pulse periodically interposed. It is believed that using pulse plating, the mass transport conditions are improved inside the features, due to adsorbtion/desorption phenomena which otherwise are not possible [17], resulting in improved fill performance, as shown in Figure 9.

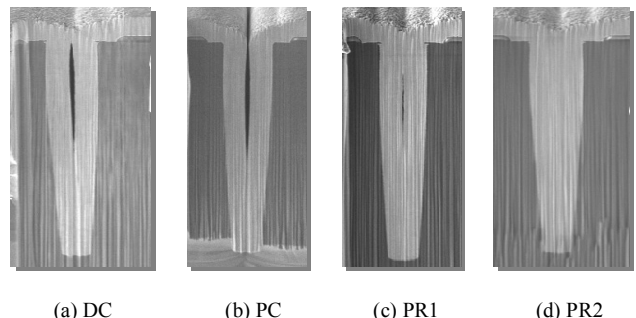

(a) DC  (b) PC  (c) PR1  (d) PR2
**Fig. 9.** Effect of waveform on via filling: (a) direct current (DC); (b) pulse current (PC); (c) and (d) pulse reverse (PR) with different reverse currents

To determine the effect of bath composition, two different electrolytes as well as variations of each of these electrolytes were tested. The results showed that bath composition as well as stability of the bath could have a significant effect of fill performance.

## VII. CONCLUSIONS

Feasibility of the copper electroplating process of forming TSV electrodes has been proven. Although the via sizes for TSV applications can vary significantly, by proper optimization of chemistries in combination with reactor design and process parameters, successful via filling can be achieved in a wide range of feature sizes and aspect ratios, as shown in Figure 10.

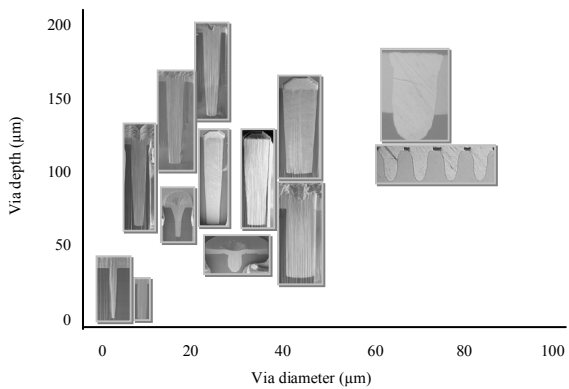

Fig. 10. Void-free electrodeposition of various TSV structures

Due to the complexity of this process and the large number of variables that could affect the final process performance, the plating mechanism is still not fully understood and continues to be investigated. In order to increase throughput and reduce the costs associated with this process, continuous efforts to develop faster and more stable processes are in progress.

## ACKNOWLEDGMENT

The authors would like to thank James Burnham, Paul Kusler and Ross Kulzer for their continuous help with development and FIB/SEM analysis and acknowledge the contribution of Bioh Kim (EVG Group) who initiated this work while employed at Semitool.